\documentclass[journal=jctc,manuscript=article,layout=twocolumn]{achemso}
\usepackage{multirow}
\usepackage{chemformula} 
\usepackage[T1]{fontenc} 
\usepackage{hyperref}
\usepackage{comment}
\usepackage[version=4]{mhchem}

\author{M. L. Pereira, Jr}
\affiliation{University of Bras\'{i}lia, College of Technology, Department of Electrical Engineering, Bras\'{i}lia-DF, 70910-900, Brazil.}
\alsoaffiliation{Rice University, Department of Materials Science and NanoEngineering, Houston-TX, 77005, United States}

\author{M. G. E. da Luz}
\affiliation{Departamento de Física \&
Multidisciplinary Laboratory for Modeling and Analysis
of Data in Complex Systems (MADComplex), Núcleo de 
Modelagem e Computação Científica-Centro Interdisciplinar
de Ciência, Tecnologia e Inovação,
Universidade Federal do Paran\'a, 81531-980 Curitiba-PR, 
Brazil}

\author{P. Cesana}
\affiliation{Institute of Mathematics for Industry, Kyushu University, 744 Motooka, Fukuoka 819-0395, Japan}

\author{A. L. da Rosa}
\affiliation{Institute of Physics, Federal University of Goi\'as, Campus Samambaia, 74690-900, Goi\^ania, Goiás, Brazil.}

\author{M. J. Piotrowski}
\affiliation[UFPel]{Department of Physics, Federal University of Pelotas, PO Box 354, Pelotas, RS, $96010-900$, Brazil}

\author{D. Guedes-Sobrinho}
\affiliation[UFPR]{Chemistry Department, Federal University of Paran\'a, CEP $81531-980$, Curitiba, Brazil}

\author{T. A. S. Pereira}
\affiliation{Physics Graduate Program, Institute of Physics, Federal University of Mato Grosso, Cuiabá, MT, Brazil}
\alsoaffiliation{Instituto de F\'isica, Universidade Federal Fluminense, 24210-346, Niterói, Rj, Brasil}
\alsoaffiliation{National Institute of Science and Technology on Materials Informatics, Campinas, Brazil}

\author{E. A. Moujaes}
\affiliation[IF]{Institute of Physics, Federal University of Bahia, Campus Ondina, 40170-115, Salvador, BA, Brazil.}

\author{A. C. Dias}
\affiliation[CIF]{Institute of Physics and International Center of Physics, University of Bras{\'{i}}lia, 70919-970, Bras{\'{i}}lia, DF, Brazil.}

\author{R. M. Tromer}
\affiliation{University of Bras\'{i}lia, Institute of Physics, Bras\'{i}lia-DF, 70910-900, Brazil}
\email{raphael.tromer@unb.br}

\title[]
{
Chalcogen Impurity Barriers in 2D Systems via 
Semi-Empirical/Machine Learning Modeling: 
A Survey over 4000 Materials
}

\begin{document}
  




\begin{abstract}
Adequate characterization of two-dimensional (2D) materials
with low energy barriers for impurity adsorption is key 
for advancing applications based on catalysis, sensing,
and surface functionalization. 
However, first-principles methods, such as Density 
Functional Theory (DFT), are often computationally extremely
expensive for feasible large-scale screenings.
Given such a scenario, we address a data-driven approach 
which integrates the semi-empirical Extended Hückel 
Method (EHM) with machine learning (ML) techniques 
to estimate adsorption energy barriers in the case of
three relevant chalcogen impurities, sulfur (S), 
selenium (Se) and tellurium (Te).
With this aim, we consider the 4036 2D materials found 
in the Computational 2D Materials Database (C2DB). 
The scheme employs the EHM to compute energy profiles 
along three in-plane migration paths, from which average
barriers can be derived. 
The equilibrium distance between the impurity and
the 2D surface is not calculated from a time-consuming
geometry optimization. 
Instead, it is estimated from a simple effective  
phenomenological expression.
Physicochemical descriptors are then obtained from the 
Matminer (Materials Date Mining) library for curated 
features. 
Four different ML models are tested, with the XGBoost 
(considering hyperparameter optimization via Optuna)
leading to the highest performance.
We further use SHAP to verify the resulting predictions, 
focusing on the $\sim1,500$ materials displaying
the lowest barrier values.
As it could be anticipated, we establish that the
average valence electron count, 
electronegativity, and atomic number are typically the 
most relevant attributes to validate the ML model. 
But we also are able to determine, for the different
chalcogen atoms, which other few descriptors likewise
considerably influence the adsorption properties.
Our results show that when combined with interpretable 
ML protocols, EHM (and potentially 
semi-empirical calculations in general) can produce a scalable framework for choosing 2D structures that 
exhibit the desired capture/release dynamics
pertinent in a variety of utilization.
\end{abstract}

\section{Introduction}

Two-dimensional (2D) materials have emerged as one of the
most promising classes of systems for advanced technological
applications, owing to their unique physical and chemical properties. 
In special, transition metal dichalcogenides 
(TMDs) --- examples being, molybdenum disulfide (MoS$_2$), 
diselenide (MoSe$_2$) and ditelluride (MoTe$_2$) ---
exhibit outstanding electronic, optical, mechanical, and 
catalytic characteristics \cite{PhysRevLett.105.136805,Chhowalla2013,Wang2012,Manzeli2017}.
Such versatility positions TMDs as attractive candidates
for various usages, including flexible electronics, 
high-performance sensors, optoelectronic devices, 
photocatalysis, and emerging energy storage technologies.

Among the distinct properties of these 2D layered 
systems, the energy barrier arising from the 
interaction with atoms or molecules is particularly 
crucial.
Indeed, this plays a central role in several
physicochemical processes, such as molecular adsorption \cite{Zhao2019}, charge transfer \cite{Dou2020}, surface 
catalysis \cite{Avasarala2023}, and chemical sensing \cite{Sahithi2020}.
Hence, when focusing on particular utilization, 
assessment of appropriate structures with 
low energy barrier values across a wide range of
TMDs may be essential \cite{han-2015}.

Density Functional Theory (DFT) stands out as one
of the primary calculation tools for accurately 
obtaining nanomaterials' structural and electronic 
properties. 
However, DFT's high computational cost significantly
limits its systematic application to the evaluation of
energy barriers over big material datasets. 
\cite{Zhao2005,Spiekermann2022}. 
This limitation poses a significant obstacle to a 
large-scale exploration of promising systems, such 
as those in the Computational 2D Materials Database 
(C2DB).
Actually, the C2DB comprises thousands of 2D compounds, 
most of them still largely unexplored \cite{Gjerding2021}.

Given such a scenario, a possible alternative is 
then to employ semi-empirical methods.
In this direction, the Extended Hückel Method (EHM) 
provides a low-cost computational approach for 
large-scale energy estimations 
\cite{Hoffmann1963,Ammeter1978}. 
Although less accurate than DFT (due to the lack of 
full geometry optimizations and the absence of explicit 
band structure inputs), the EHM is highly useful in
screening total energies of big pools of nanostructures
\cite{Farrar2022}, eventually as a first selective
procedure for the best candidates in practical times.
For example, such a strategy has already been shown to 
work well in predicting the adsorption energy of 
contaminants on 2D surfaces
\cite{Liang2016,Dysart2016,Ren2022,Tong2021,Sharma2024}, 
including the case of alkali metals (Na, K, and Li) 
with graphene \cite{Olsson2019,Heidari2023,Ullah2019}.

In parallel, machine learning (ML) techniques have 
revolutionized the study of material features, 
enabling the efficient prediction of complex physical
quantities from a limited number of reference calculations \cite{Butler2018}. 
Moreover, the application of ML in high-throughput 
probing should facilitate trend identification and 
accelerate the uncovering of materials with 
tailored properties. 
Certainly, a critical step in the scheme is the
construction of representative descriptors for the
systems under scrutiny.
But a robust tool to address such difficulty is the 
\textit{Matminer} (Materials Data Mining) library,
skillfully generating effective quantitative 
representations for ML models \cite{Ward2018}.

\begin{figure*}[!t]
    \centering
    \includegraphics[width=\linewidth]{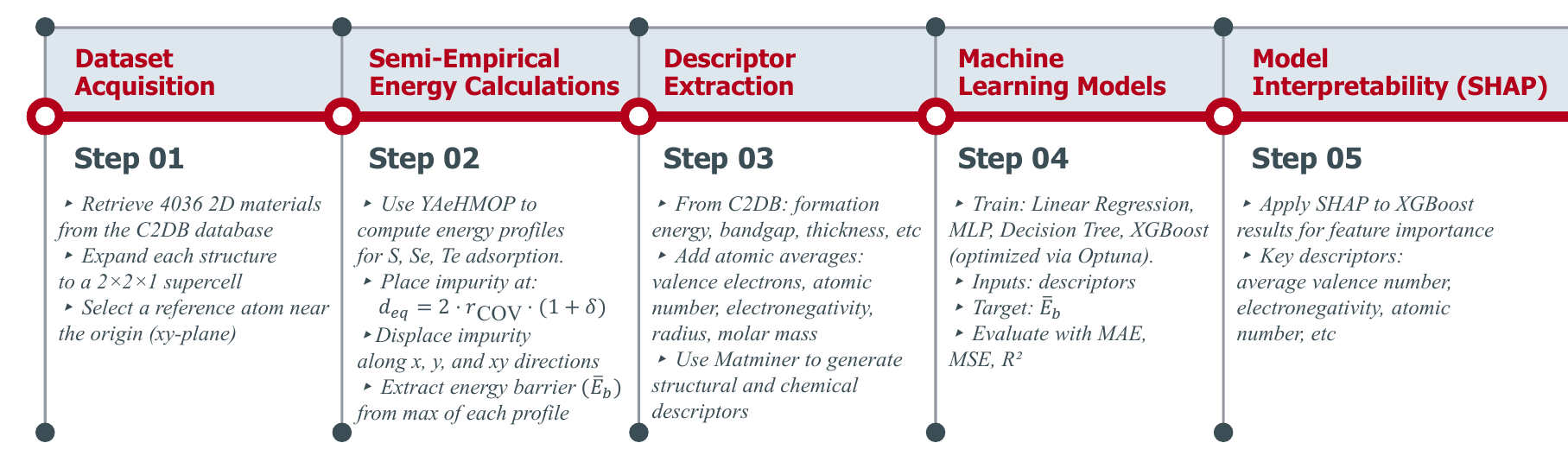}
    \caption{Workflow summarizing all the data gathering,
    calculations, ML models and interpretative analyzes implemented 
    to properly characterize the adsorption of chalcogen 
    atoms in 2D materials classified
    in the C2DB database (comprising 4000+ distinct systems).}
    \label{fig:figure1}
\end{figure*}

Among the many available supervised learning 
algorithms,  XGBoost (Extreme Gradient Boosting) has
emerged as a high-performance protocol for handling
heterogeneous elaborated datasets 
\cite{10.1145/2939672.2939785,Zhang2018}. 
Further, to enhance model interpretability, 
techniques such as SHAP (SHapley Additive exPlanations)
have been increasingly considered, allowing for the
quantification of descriptors importance and 
for the inference of actual physical correlations.
This establishes explicit connections between the most 
relevant features and the underlying physicochemical 
mechanisms governing the system predicted properties \cite{NIPS2017_8a20a862,Rodríguez-Pérez2020}.

Here we develop a novel framework to investigate 
energy barriers in 2D structures by integrating EHM 
with advanced ML techniques.
We then address chalcogen atoms, specifically,
Sulfur (S), Selenium (Se), and Tellurium (Te),
interacting with 2D materials surfaces.
We take into account the problem geometry through 
a phenomenological, nevertheless rather simple and
effective, expression based on the covalent radii 
of the adsorbates.
It gives the fixed equilibrium distance between the 
atom and the 2D surface.
From the proposed methodology, we estimate the 
energy barriers for 
over 4,000 systems retrieved from the C2DB database. 
The obtained large dataset is thus used to train 
predictive models (in different ranges of energy)
employing the XGBoost --- for sake of comparison,
other ML procedures are also tested.
Finally, analyzes considering the SHAP protocol 
for the 1,495 materials with the lowest barriers
lead to accurate predictions and tangible physical 
interpretation of the factors governing energy 
barriers in these structures.
The generality of the whole procedure might 
constitute a relevant new approach for studying 
distinct aspects of atoms deposition on 2D 
materials.

\section{Methodology}
\label{sec:methodology}

In the present study we follow a sequential protocol 
that involves (see the workflow in Fig. 
\ref{fig:figure1}):
(i) to built an energy barrier dataset (steps 01 and 02); 
(ii) to extract physicochemical descriptors for 
numerical representation of each two-dimensional 
system (step 03); and
(iii) to implement ML models based on
the physicochemical descriptors (step 04)
and then to physically interpret the obtained results 
(step 05).

For the actual systems, we accessed one of the most 
comprehensive repository of 2D structures, the C2DB \cite{Gjerding2021}, covering a rich diversity of 
configurations, symmetries, and fundamental 
physicochemical properties. 
It comprises an extensive collection of 2D 
materials, from experimentally synthesized 
ones, such as graphene, to compounds proposed
either theoretically or identified through AI-assisted 
simulations.

Our framework steps are described below.

\begin{figure}[t!]
    \centering
    \includegraphics[width=\linewidth]{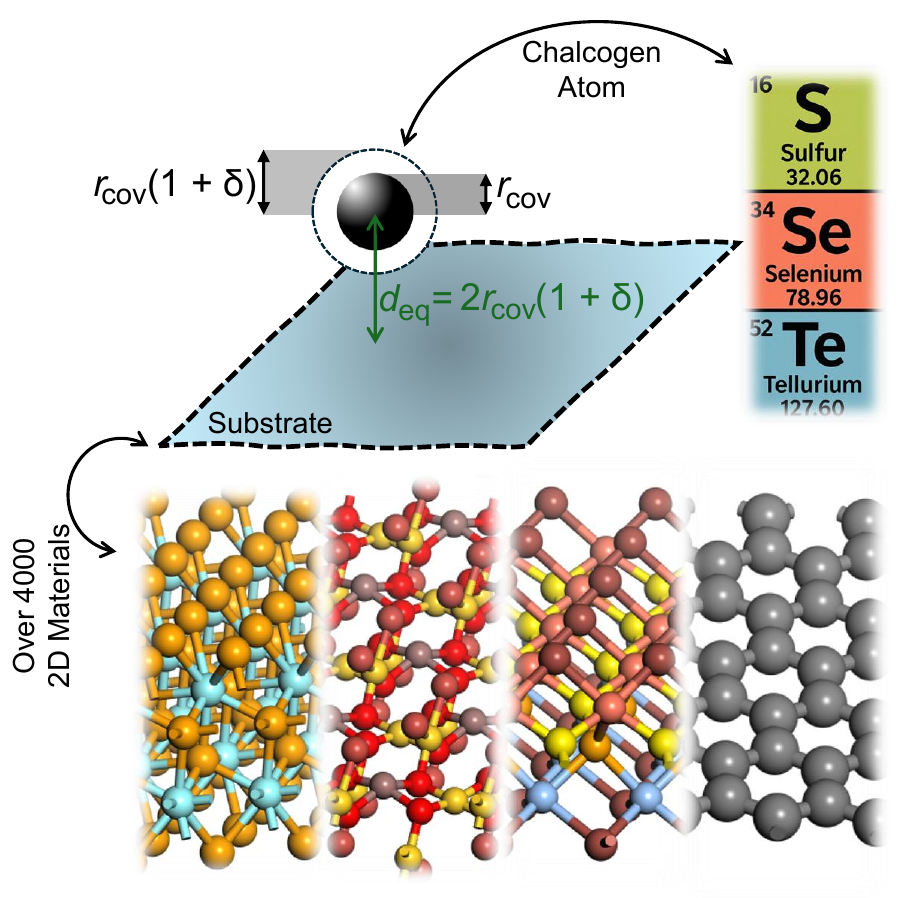}
    \caption{Schematics of the problem geometry.
    Distinct 2D material (from the C2DB database) 
    constituting the substrate for a chalcogen 
    atom (either S, Se or Te) adsorption.
    The effective equilibrium distance $d_{eq}$, 
    used to calculate the energy barriers, is estimated 
    from the phenomenological considerations in Sec. 
    \ref{d-heuristic}.
    }
    \label{fig:figure2}
\end{figure}

\subsection{Semi-empirical calculations}
\label{sec:2-1}

For each structure available in the C2DB, we estimated
the energy barrier associated with the adsorption of three
types of dichalcogenide impurities, sulfur (S), selenium 
(Se), and tellurium (Te), by employing the EHM, as in 
the YAeHMOP software package \cite{avery2018extended}. 
This semi-empirical method can calculate the system
total energy, but without performing any geometry 
optimization, which requires a predefined equilibrium 
distance $d_\text{eq}$ between the impurity and the 2D 
surface; refer to Fig. \ref{fig:figure2}.
This $d_\text{eq}$ is thus estimated from the
scheme proposed in Sec. \ref{d-heuristic}.

Each 2D material was expanded into a $2\times 2\times 1$ 
supercell, and a reference atom was selected as the one closest to the origin within the $xy$ plane. The impurity was initially placed above this reference atom at the equilibrium distance $d_\text{eq}$. Subsequently, the impurity was displaced along three distinct in-plane trajectories: the $x$ axis, the $y$ axis, and the $xy$ diagonal. Each trajectory spanned a displacement of approximately 3 \r{A}, beyond which edge effects
--- due to the cluster approximation adopted in YAeHMOP for periodic boundary conditions ---
began to introduce non-physical artifacts in the energy profile. Therefore, three energy barrier curves were generated for each system, corresponding to the three directions. The average energy barrier was computed as
\begin{equation}
\bar{E}_b=\frac13\left(E_b^x+E_b^y+E_b^{xy}\right),
\label{eq:energy-average}
\end{equation}
where the maximum value from each energy profile was selected as the target variable for ML models.

\subsubsection{The semi-empirical equilibrium distance}
\label{d-heuristic}

Two-dimensional materials are often characterized by
in-plane covalent bonds and out-of-plane van der Waals 
interactions \cite{liu-2023}.
The latter being responsible, for example, for the separation
distance between graphene sheets in distinct bilayer
structures (typically in the range $3.25\sim3.50$ \AA) 
\cite{fukaya-2021} and for atom adsorption
\cite{gao-2015,bian-2022}.
Thus, the van der Waals radius  \cite{batsanov-2001a} --- 
associated to an effective size for the interfacial 
adhesion mechanism --- is one of the main factors 
determining the equilibrium distance  $h_\text{eq}$ 
between the adsorbed atom and distinct points along
the 2D material surface \cite{rokni-2020}.
Moreover, some findings in the literature
\cite{duffy-2016,cariglia-2018} have pointed out
that atom adsorption in 2D materials tend to display
rather simple universal relations to the adsorbent
structure geometrical symmetries.
Hence, in a first order approximation, where we 
suppose an effective equilibrium distance $d_\text{eq}$
for the full surface, such a configuration parameter 
can be assumed fairly independent on specific details 
of the atoms and 2D materials involved
(but see additional discussion in section 
\ref{sec:remarks}).

Based on the aforementioned, we propose to set
$d_\text{eq} = 2 \, r_\text{vdW}$, with
$r_\text{vdW}$ the van der Waals radius.
Nonetheless, it has been long known
\cite{batsanov-2001a,batsanov-2001} that for a large
number of situations $r_\text{vdW} - r_\text{cov} = 
\Delta_r > 0$, for $r_\text{cov}$ the atom covalent 
radius and $\Delta_r$ reasonably constant. 
So, $d_\text{eq}$ can be written as
(see the schematics in Fig.  \ref{fig:figure2})
\begin{equation}
d_\text{eq}  = 2 \, r_\text{cov} \, (1 + \delta),
\label{eq:d}
\end{equation} 
where $\delta = \Delta_r/r_\text{cov}$.

For weak van der Waals bonding (like in gases,
liquids and polymers), the Pauling rule proposes that
$\Delta_r|_{\mbox{\scriptsize Pauling}} = 0.8$ \AA, a 
value put in solid grounds in ref. \cite{batsanov-2001}.
For the clearly tighter attraction between adsorbed
atoms and 2D materials \cite{gao-2015,bian-2022}, 
such $\Delta_r$ must be smaller.
So, based on some estimations in the literature
(see, e.g., scheme 5 and Sec. 5.4 of ref. 
\cite{alvarez-2013}) we assume $\Delta_r \sim 
\Delta_r|_{\mbox{\scriptsize Pauling}}/2 = 
0.4$ \AA.
Given that the three atoms of interest here are
S, Se and Te of $r_\text{cov}$, respectively,
equal to 1.02 \AA, 1.2 \AA \ and 1.36 \AA, we get
$\delta \approx 0.3$.
So, Eq. (\ref{eq:d}) with $\delta=0.3$ is the
expression we consider throughout this
work.

We remark that as a further support for the 
approximation in Eq. (\ref{eq:d}), it properly 
reproduces average $h_\text{eq}$'s from DFT 
calculations for alkali metal atoms 
(Na, K, Li) interacting with graphene and other 
2D surfaces
\cite{Olsson2019,Heidari2023,Ullah2019}.

\subsection{The physicochemical descriptors}
\label{sec:2-2}

To numerically characterize each system (see
Fig. \ref{fig:figure1}, step 03), we considered
a set of physicochemical descriptors derived directly 
from the C2DB database; as examples we cite 
formation energy,  monolayer thickness, and 
electronic band gap. 
These were complemented by externally computed attributes,
including average: valence electrons, atomic number, 
atomic radius, electronegativity and molar mass. 
In addition, we employed the Matminer library \cite{Ward2018}, 
which provides a broad range of automated descriptors 
based on chemical composition, crystal structure, and 
electronic features.
Matminer has been widely adopted in materials 
informatics
due to its ability to generate robust, interpretable, 
and reproducible properties for usage in ML
pipelines \cite{Li2024}.

The totality of these features, namely, those 
directly from the C2DB, plus the computed 
attributes, plus those from the Matminer, we 
call the Full Set of  Descriptors, FSD.

\subsection{The ML and interpretability
models considered}

The generated and assembled dataset of descriptors,
the FSD, was considered to evaluate and compare the 
performance of four supervised learning algorithms.
The first was a simple model that assumes a linear 
relationship between the descriptors and the target 
energy barrier. 
Although interpretable and straightforward, linear
regression is naturally
insufficient for systems with considerable nonlinear 
trends \cite{kuhn2013applied}. 
Next, we employed a multilayer perceptron (MLP) neural 
network, capable of capturing complex polynomial patterns,
nonetheless requiring great amount of data and 
hyperparameter tuning \cite{Schmidhuber2015}. 
As a third possibility, we tested a decision tree model,
which works through hierarchical decision rules by 
partitioning the descriptor space.
While highly interpretable, this approach is susceptible 
to overfitting noisy data \cite{Kingsford2008}. 

Lastly, we assumed the XGBoost algorithm, a powerful
ensemble method based on gradient-optimized decision 
trees. 
XGBoost is greatly recognized for its highly predictive
accuracy, resilience to overfitting, and excellent 
performance across large and heterogeneous datasets. 
Indeed, in our analyzes this scheme outperformed 
all the other three protocols \cite{10.1145/2939672.2939785}.

To interpret the outputs of the ML models and to 
identify the most influential features governing 
energy barrier predictions, we used the SHAP method \cite{NIPS2017_8a20a862}. 
Based on cooperative game theory, SHAP assigns a 
Shapley value to each input feature, quantifying its 
marginal contribution to the model's prediction.
SHAP has lately earned considerable popularity in 
materials science.
It allows to give concrete interpretations
to results obtained from complex black-box models. 
In other words, SHAP helps to link physical
attributes to forecast behavior from AI-based
methodologies.
This is fundamental to ascribe phenomenological
significance to context-free procedures, especially
if one is trying to unveil the physical-chemical 
mechanisms underlying the properties of a material
\cite{Choudhary2022}.

\section{Results}

\subsection{A relevant benchmark: \\ 
graphene as substrate}

We start our analysis supposing graphene as the 2D
substrate for the S, Se and Te dichalcogenide 
elements, given the graphene well-known 
physical and chemical properties and its widespread 
usage as a reference in surface science.
So, this is an ideal case (a benchmark) to validate 
the consistency and reliability of the proposed 
semi-empirical methodology. 

\begin{figure}[!t]
\centering
 \includegraphics[width=1.25\linewidth]{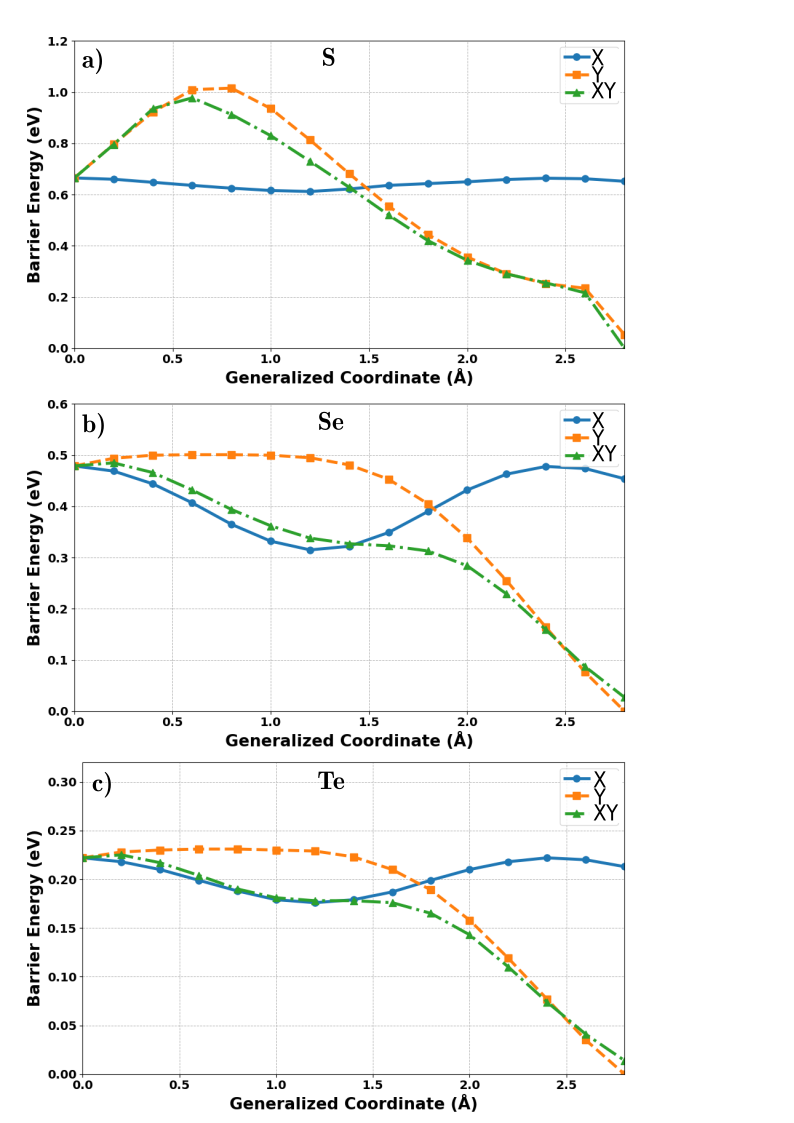}
\caption{Energy barrier profiles (in eV) obtained for the dichalcogenide elements: (a) S, (b) Se, and (c) Te, 
adsorbed on graphene using the EHM via the YAeHMOP 
software. 
In each case, the impurity was displaced along three 
directions over the graphene surface: $x$ (blue), 
$y$ (orange), and the diagonal $xy$ (green).}
\label{fig:fig1}
\end{figure}

We present the energy barrier profiles obtained for
S, Se and Te, respectively, in Fig. \ref{fig:fig1}(a), \ref{fig:fig1}(b) and \ref{fig:fig1}(c). 
Each graph displays the variation in the energy barrier as 
function of a generalized reaction coordinate, 
corresponding to three distinct in-plane diffusion paths 
for the impurity along the: $x$ (blue), $y$ (orange), 
the diagonal $xy$ (green), directions. 
These trajectories simulate the lateral migration of the
adsorbate along the 2D surface, allowing to characterize
directional effects on diffusion. 
We recall that as discussed in Sec. \ref{sec:2-1}, the 
displacements were limited to approximately 3~\r{A},
a constraint introduced to reduce edge effects
from the simplified periodic boundary condition 
treated via cluster approximation by YAeHMOP 
software.

As clear from Fig.~\ref{fig:fig1}(a), the energy 
barrier profile for S shows marked anisotropy among 
the $x$ and $y$ displacement directions.
Indeed, the highest barrier takes place along the $y$ 
direction, peaking at approximately 1.02~eV, whereas
the lowest barrier lies along the $x$ direction, reaching
only 0.66~eV --- but for a profile which is almost
constant.
Mostly influenced by $y$ (in fact, closely following 
it), the $xy$ path has a peak at 0.98~eV. 
This pronounced directional dependence seems to be
be attributed to the graphene surface's local electronic
topology and the S atom's preferential adsorption 
orientation. 
Notably, anisotropic interactions have been reported 
in DFT-based computations \cite{Stark2019,Hou2022}, 
arising from variations in the registry between the 
impurity and the graphene lattice.
Energy barrier values for S on various 
surfaces typically range from 0.5 to 1.2~eV, depending 
on the adsorption site and surface coverage \cite{BernardRodríguez2018}. 
Our much simpler semi-empirical calculations fall
within this expected range.

Results for Se are depicted in Fig.~\ref{fig:fig1}(b).
Note that overall it displays a shorter range of
energy variation than S. 
The $y$ direction continues to have the largest 
barrier, reaching a maximum of roughly 0.50~eV;
the $x$ and $xy$ directions show maximums that are
only somewhat lower than it.
Also, differently from S, for Se the $xy$ path is
now a more balanced combination of the behaviors 
along $x$ and $y$.
This decrease in magnitude and anisotropy might be 
attributed to the larger atomic radius and lower 
electronegativity of Se relative to S, 
weakening the orbital overlap with the graphene
surface. 
Thus, the potential energy landscape becomes 
less corrugated, with shallower energy minima and maxima. 

\begin{figure}[t!]
\centering
 \includegraphics[width=0.95\linewidth]{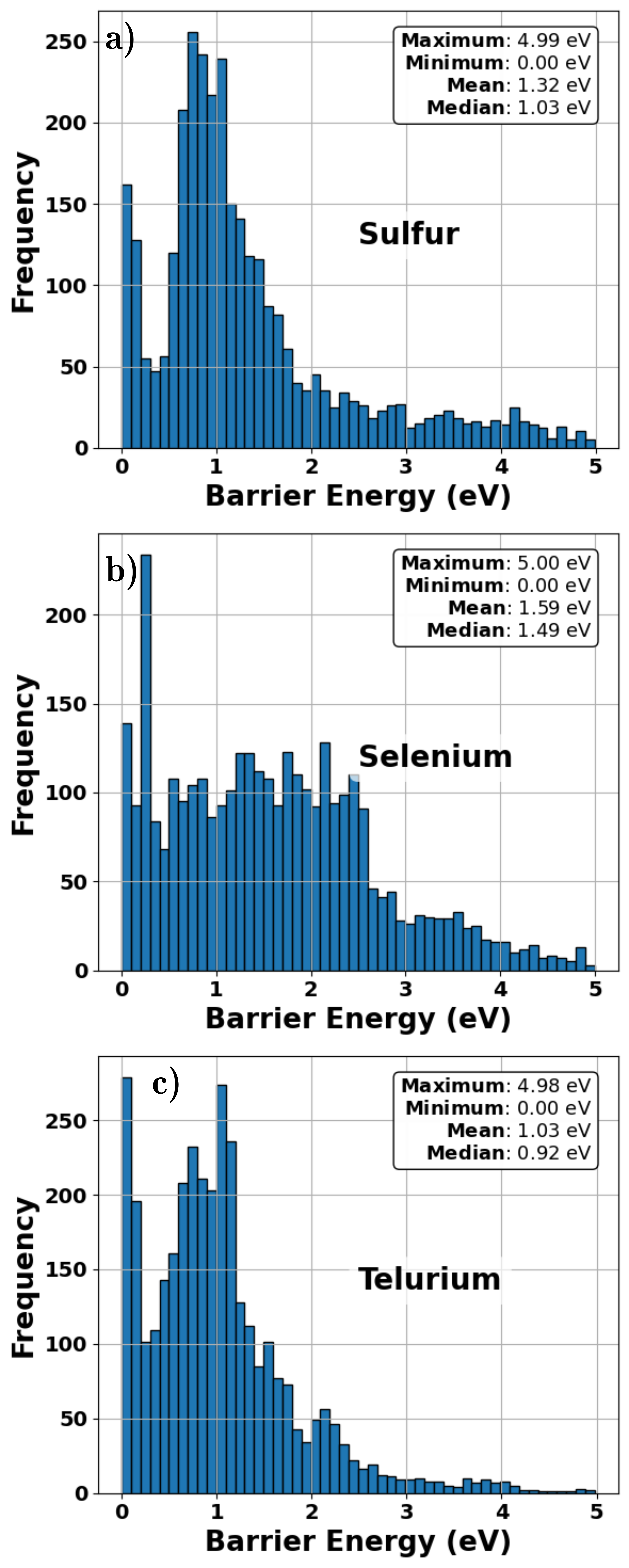}
\caption{Histograms of average energy barriers 
for about 3150 2D systems from the C2DB database,
for which the energies do not exceed 5.0 eV.
The impurities are (a) S, (b) Se and (c) Te. 
The values represent the average over the three 
directions $x$, $y$, and $xy$.}
\label{fig:fig2}
\end{figure}

These trends are consistent with theoretical studies
of Se adsorption in graphene and related 2D materials,
where typical barriers of $\sim0.6$~eV are often obtained 
\cite{Hou2022}. 
This enhanced surface mobility of Se can have 
practical implications for surface functionalization, epitaxial growth, and thermally activated diffusion. 
Hence, further characterization through ML 
approaches may provide valuable insights 
into design and selection of functional materials
(see later).

The barrier energies for Te are displayed in 
Fig.~\ref{fig:fig1}(c), exhibiting the lowest values
among the three elements, although with general
traits resembling Se. 
The maximum is approximately 0.23~eV for $y$ 
and about 0.22~eV for the $x$ and $xy$ directions.
The Te quasi-isotropic behavior (in terms of
maximum barriers heights) reflects its larger 
polarizability and atomic radius, which lead to
weaker and 
more delocalized interactions with the graphene
surface. 
Consequently, the energy landscape becomes flatter
and less sensitive to the diffusion direction. 
All this aligns with previous investigations of 
Te adsorption on graphene, yielding barriers usually 
below 0.3~eV \cite{Nascimento2020}. 
Thus, the EHM properly captures the correct
interaction energies relative magnitudes and key
symmetry trends; critical features for the 
understanding of surface diffusion and transport
phenomena.

An important qualitative point to be emphasized is 
that, despite the simplicity of the semi-empirical 
scheme used, particularly its lack of geometry 
optimization and electronic structure resolution, 
the order of the calculated energy barriers (S > Se 
> Te) concurs with the expected chemical tendency 
across the periodic table group 16:
the increase in atomic radius and the decrease in electronegativity as one moves down the 
chalcogen group.
This results in progressively weaker 
effective interactions between the impurity and the 
substrate.

\subsection{Energy barrier computations for the full
C2DB database}
\label{sec:3-2}

Having illustrated the reliability of the present
approach for the paradigmatic graphene case, next 
we apply our semi-empirical method to a rather large collection of systems.
Indeed, for the 4036 2D materials from the
C2DB database, each surface is doped with one of three 
dichalcogenide impurities.
Then, the average energy barrier is calculated 
from the maximum energy barrier along the $x$, $y$, 
and $xy$ directions; refer to Eq. 
(\ref{eq:energy-average}).
In Fig.~\ref{fig:fig2} we summarize the simulations 
by means of histogram distributions (i.e., frequency 
or number of cases within a given energy interval, 
the bin).
To simplify the analysis, we show only the materials
with barrier energies up to 5.0 eV.
This reduces the number of materials to about 3150 
(78\% of the original set) for each impurity atom, 
namely, S (Fig.~\ref{fig:fig2}(a)), Se
(Fig.~\ref{fig:fig2}(b)), and Te
(Fig.~\ref{fig:fig2}(c)).

We should emphasize that the C2DB comprises both 
already synthesized as well as theoretically predicted 
2D systems.
Moreover, it covers many distinct structural families.
In this way, the present survey characterizes the 
energy barriers behavior across a great variety  
of 2D materials.
On one hand, it pinpoints the variations in 
substrate interactions, influenced by structure, 
symmetry, and electronic properties.
On the other hand, it unveils the differences in the 
chemical nature of the dichalcogenides adsorbing.
Therefore, the histogram reveals certain coincidences
as well as important distinctions between the three
atoms.

As for similarities, for the energy interval 
considered, 0$\sim$5 eV, the majority of the examples
are below 2.0 eV.
Also, there is a considerable number of materials for
which the barriers are in the range 0$\sim$0.3 eV.
The mean and median for S, Se and Te are,
respectively, 1.32 and 1.03 eV, 1.59 and 1.49 eV
and 1.03 and 0.92 eV.
So, the difference between the highest (Se) and
lowest (Te) mean, of about 35\%, is not so large. 

Regarding the contrasts among S, Se and Te, we 
first observe that the histogram for S,
Fig.~\ref{fig:fig2}(a), demonstrates a certain
similarity among materials, with the principal 
mode being around 0.6$\sim$1.6 eV.
Then, there is a long tail extending toward higher 
energies (up to 5 eV).
Consequently, the distribution exhibits a considerable
skewness.
Further, it shows that although most systems present
moderate barriers, a significant number of surfaces 
have a strong interaction with S.
This is likely due to the Sulfur high propensity to
form covalent bonds with the 2D materials orbitals
of high electronic density.
Such type of orbitals are particularly common in 
transition-metal-based structures.

In distinction, the Se histogram in
Fig.~\ref{fig:fig2}(b) displays a broad plateau,
with fairly similar frequencies in the energy range 
0.6$\sim$2.5 eV. 
The average for Se is 1.59 eV, while for S it 
is 1.32 eV.
Such a higher value can be attributed to a mildly 
lower electronegativity of Se, resulting in a more
diffuse interaction with the surfaces, but which 
is still enough to yield relevant barrier values. 
Likewise, the tendency of a stronger orbital 
polarization for Se, that depends on the lattices
symmetries --- i.e., hexagonal, orthorhombic,
etc --- may also contribute to 
a wider range for the energy barrier values.

Overall, the histograms for S and Te, respectively,
Fig.~\ref{fig:fig2}(a) and Fig.~\ref{fig:fig2}(c),
display certain parallels, however differences are 
also identified.
For instance, the distribution for Te exhibits a 
faster decreasing tale.
Moreover, the number of surfaces having barrier 
energies in the range 0$\sim$0.3 eV for S is only 
of around 60\% of those for Te.
These facts lead to a mean of 1.03 eV for Te (the 
lowest among the three impurities), suggesting that
generally Te interacts weakly with the 2D systems, 
the same trend found for graphene as a surface
in the previous section.
This should be expected, considering the Te larger
atomic radius, lower electronegativity, and higher
polarizability, when compared to S and Se.
It is worth remarking that from DFT-based computations, nonetheless implemented for much fewer situations \cite{Nascimento2020}, it has been demonstrated the
somehow feeble binding nature of Te.
Here, from our relatively inexpensive protocol, we 
have been able to extend such predictions for a 
rather big set of 2D materials.

\begin{figure*}[htb!]
\centering
 \includegraphics[width=0.86\linewidth]{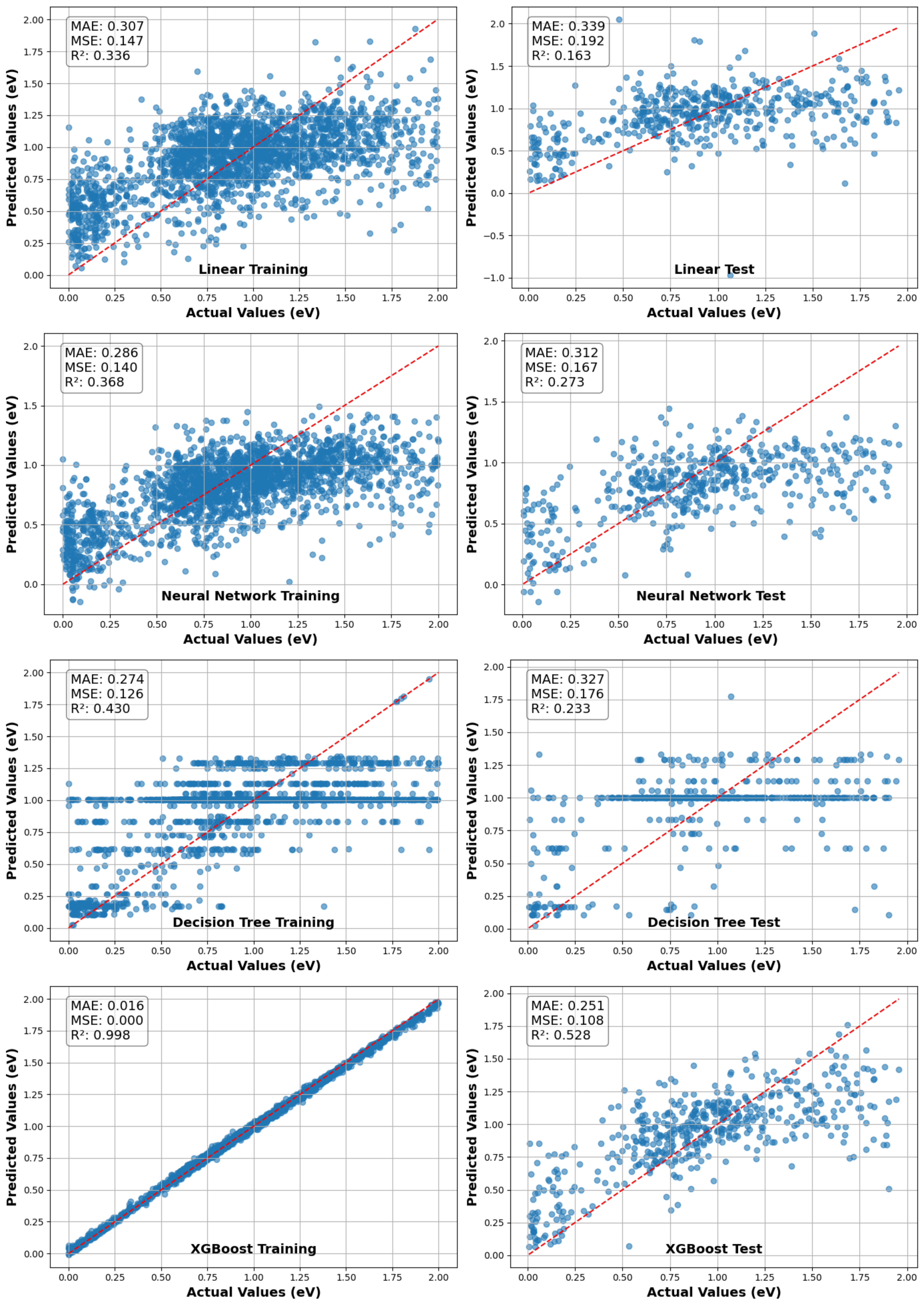}
\caption{Comparison between the calculated and ML-predicted energy barrier values 
($\leq 2.0$ eV) for the S impurity adsorbed on materials 
of the C2DB database.
The ML models are: 
Linear Regression (first row), Neural Network (second row), 
Decision Tree (third row), and XGBoost (fourth row). 
Each model training stage is shown on the left and the 
testing stage on the right. 
The dashed red line represents the ideal $y=x$ identity line. Performance metrics, namely, Mean Absolute Error, Mean 
Squared Error, and the coefficient of determination $R^2$, 
are also shown.}
\label{fig:fig3}
\end{figure*}

\subsection{ML models predictions for the barrier energies}

In critical applications, such as those related to 
catalysis and building functionalized surfaces with 
adjustable diffusion characteristics, adsorbed atoms 
should exhibit high surface mobility, which translates 
into low energy barriers.
Hence, for the ML analyzes here we further restrict
the materials to those with energy barriers up to
2.0 eV.
This amounts to around 2560 2D systems from the 
original 4036, i.e., 63\% of the total of materials
in the C2DB database.

Although semi-empirical calculations tend to
display certain differences from the barriers actual
values, the global statistical patterns observed in
the distributions like, spread, shape, mode 
position, and average energetic relative order 
(Se > S > Te), comply with known physicochemical 
traits and with much more sophisticated theoretical 
and experimental analyses in the literature.
This supports the use of our obtained results
as reliable input data \cite{mohammed-2025} to 
implement ML models.
We also recall that the predictors considered for the
ML models below (and for the SHAP study in Sec.
\ref{sec:3-4}) are all those mentioned in Sec. 
\ref{sec:2-2}, the FSD.

For the ML outcomes presented below, we stress that
the parameters of each algorithm were thoroughly 
tested and tuned to optimize their predictive 
performance.
More concretely, for simpler models, like linear 
regression and decision trees, various combinations of 
standard settings were explored, including maximum tree 
depth, splitting criteria, and normalization strategies. 
Further, an MLP architecture with various hidden layers was 
considered for the neural network, and several 
configurations entailing batch size, activation functions, 
and optimization schemes were tested. 
In the case of XGBoost, which delivered the best results,
a fine-tuning procedure was carried out using the Optuna 
library \cite{10.1145/3292500.3330701}. 
It allowed an efficient hyperparameter search, targeting 
the number of estimators, learning rate, maximum depth, 
and subsampling ratio.
Such protocol significantly improved the model accuracy 
--- especially in the testing phase, where  generalization 
capability is critical.

The comparison between the ML models performances 
yields essentially the same conclusions for the 
three adsorbing atoms addressed in this work. 
So, the following discussion focuses on the S case
in detail, with only a brief summary provided for 
Se and Te.
The scatter plots in Fig.~\ref{fig:fig3} for S 
compare the ML predictions (for all the four models 
investigated) with the calculations of Sec. 
\ref{sec:3-2}.
In Fig.~\ref{fig:fig3} each row corresponds 
to one model, with the left (right) panels depicting 
the training (test) results. 
The performances are quantified through the Mean 
Absolute Error (MAE), Mean Squared Error (MSE), and 
the coefficient of determination (R$^2$), shown in 
the corresponding graphs.

For the linear regression in the top row (certainly the 
simplest ML model), as expected the performance is rather
modest, with R$^2$ equals to 0.336 and 0.163 for the 
training and testing sets, respectively.
It also leads to MAE close to 0.31 eV (training)
and a very broad dispersion --- refer to the 
diagonal dashed red line.
This stems from a linear fitting 
trying to describe the nonlinear relationship 
between the assumed descriptors and the target energy
barrier.

The second row in Fig.~\ref{fig:fig3} shows the results for
the neural network model based on an MLP architecture.
Now, $R^2$ and MAE are, respectively, 0.368 and 0.286 eV 
(training) and 0.273 and 0.312 eV (testing). 
Thus, some improvement is achieved, although still
limited.
This may be attributed to the relatively simple nature of 
the input features, based mainly on compositional averages, 
and dataset size: neural networks are notoriously dependent 
on significantly large samples, despite being of low
computational cost, particularly when coupled with
architectural and hyperparameter optimization 
\cite{Schmidhuber2015}.

The decision tree model is presented in the third row 
of Fig.~\ref{fig:fig3}.
It has the important feature  of partitioning 
the input space through a hierarchy of decision rules. 
While leading to R$^2 = 0.430$ during the training 
stage, it suffered a notable drop in performance 
during testing, with R$^2$ decreasing to 0.233 and 
the MAE increasing from 0.274 eV to 0.327 eV. 
This behavior suggests overfitting, a well-documented 
limitation of single decision trees when applied to noisy 
or highly variable data \cite{Quinlan1986}. 
Also, the layer-like pattern in the plots reflects 
the discrete nature of the model predictions (owned 
to the countable rules), resulting in a poor 
representation of any target variable inherently 
continuous, of course the case of energy barriers.

Finally, the fourth row of Fig.~\ref{fig:fig3} 
depicts the XGBoost outcomes, outperforming the 
other three models in every respect. 
During training R$^2 = 0.998$, with negligible
MAE and MSE, indicating an almost perfect fit. 
In the testing stage, XGBoost maintains the robustness,
with R$^2 = 0.528$, MAE = 0.251 eV, and MSE = 0.108.
This success can be attributed to the model ensemble 
nature, combining many weak learners (shallow 
decision trees) through gradient boosting to minimize 
prediction error. 
XGBoost also incorporates built-in regularization 
techniques, mitigating overfitting.
Such a superiority has already been reported
for predicting other material characteristics, 
e.g., band gaps, formation energies, elastic 
constants, and thermal coefficients 
\cite{10.1103/PhysRevLett.120.145301}.

Akin analyzes for Se and Te showed essentially the 
same qualitative results concerning the efficiency 
of the distinct ML models, again with XGBoost 
performing considerably better.
For instance, for Se the $R^2$ values in the test stage
were 0.151, 0.309, 0.260, and 0.520 for linear regression, 
neural network, decision tree, and XGBoost, respectively.
For Te, the corresponding values were 0.319, 0.353, 
0.313, and 0.540.
The full set of calculations for Se and Te are given 
in the Supplemental Information (SI). 

The above upshots demonstrate that while simpler
ML schemes, as linear regression and decision trees,
offer baseline insights, they lack the required 
generality and plasticity to unveil the intricate
relationships between atomic descriptors and more 
tangled quantities, the case of energy barriers. 
Even the more elaborated neural network approach yielded
limited gains.
On the contrary, XGBoost effectively displayed the proper
balance between accuracy and stability to deal with a
large and diverse sample of materials.

\begin{figure*}[htb!]
\centering
 \includegraphics[width=0.9\linewidth]{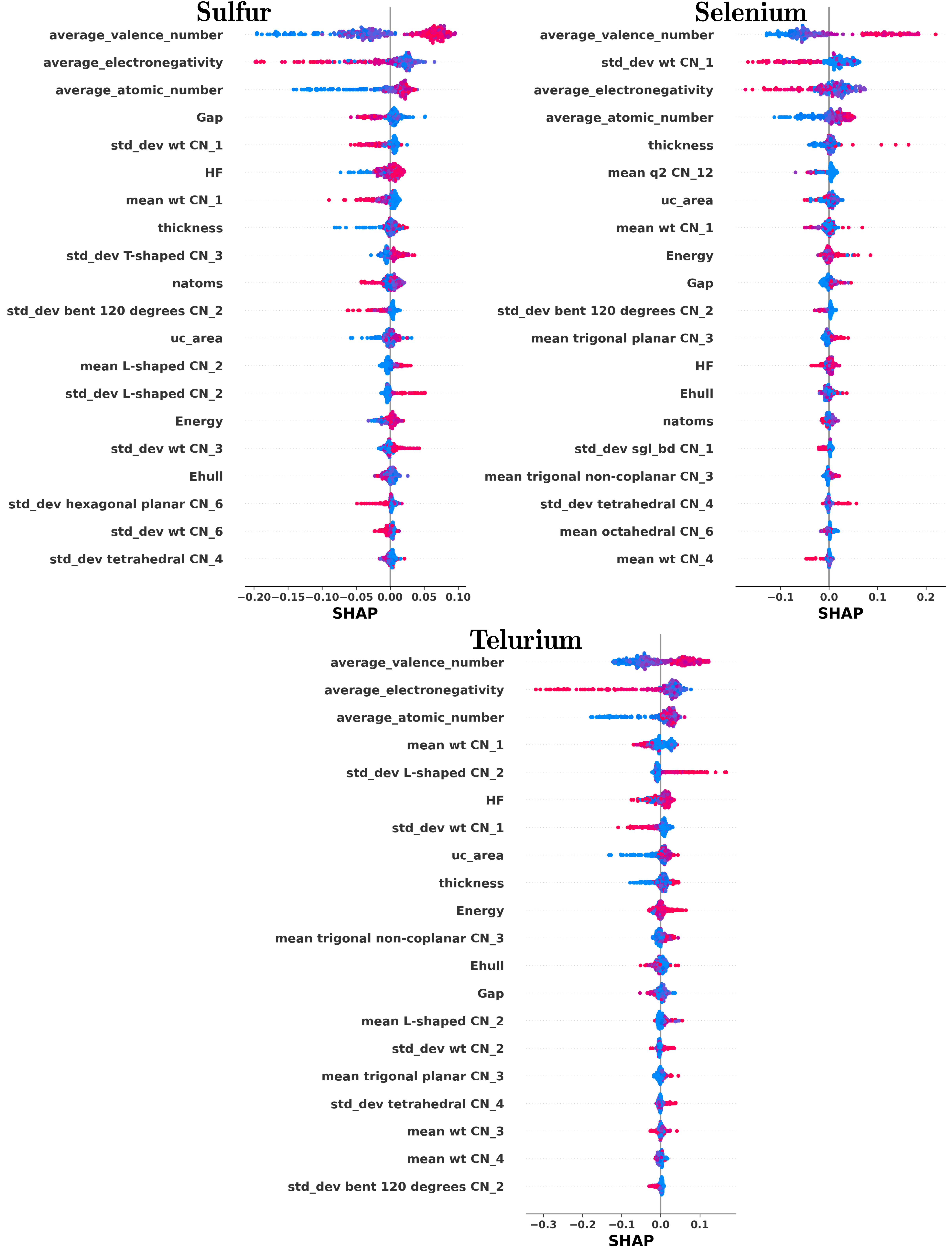}
\caption{Interpretability analysis using the SHAP 
protocol applied to the energy barrier estimations 
from the XGBoost model.
The graphs depict the results of adsorption
of either S (top left), Se (top right), or Te (bottom)
on the two-dimensional systems from the C2DB database
with barrier energy $\leq 1.0$ eV.
Each point represents a distinct 2D material.
Predictor features are ranked vertically in 
descending order of their importance in the model. 
A predictor variable value is high (low) if its color 
is pink (blue).
The horizontal axis corresponds to the SHAP index.
Positive (negative) values tend to increase (decrease)
the energy barrier.}
\label{fig:fig6}
\end{figure*}

\subsubsection{Examining the ML predictability}

Considering the best-performing model, XGBoost,
for the relatively broad barrier energy range of 0-2.0 eV, 
all chalcogen atoms exhibited strong consistency between 
cross-validation and held-out test performance. 
This indicates that the adopted hyperparameter
optimization and data partitioning strategy effectively
migrated information leakage.
However, the train–test error gap remained substantial 
($\sim$0.25--0.32 in RMSE), a clear sign of overfitting
when the task covers a wide energy range with higher 
intrinsic variance.

Repeating the XGBoost modeling, but restricting the 
set of materials to those with barrier energies 
$\leq1.5$ eV and then to those with $\leq 1.0$ eV
(in this last case the number of systems being around
1490), the predictive accuracy improved markedly: 
test RMSE dropped from $\sim$0.34 to $\sim$0.19--0.21, 
with MAE following the same trend, indicating that the 
models capture the structure–property relationship much 
more reliably for narrower ranges for the barrier
energy.
Importantly, the train–test gap contracted steadily, 
from 0.25–0.32 to nearly 0.13–0.17.
Further, at a very restrictive regime (<0.5 eV), 
roughly representing 500 2D materials, errors reached 
exceptionally low values (test RMSE ~0.09–0.11, MAE ~0.07),
while the overfitting gap decreased to ~0.07–0.10.

Regarding the ML results for the different atoms
and for the aforementioned smaller energy windows, 
Se continuously reduced absolute error and showed
the largest decrease in RMSE and MAE.
In contrast, Te emerged as the most robust and 
generalizable element, achieving the highest 
explained variance ($R^2$) and the smallest 
overfitting gap as the target range narrowed.
This suggests that for Te, the model captures a 
broader portion of the underlying  variance without
holding up spurious details. 
Finally, S remained a stable baseline with competitive 
performance and slightly smaller gaps in some
intermediate regimes, but it was generally surpassed 
by Se in raw error and by Te in variance explanation.

As a last technical summary, across all energy thresholds 
for all atoms, the hyperparameter search consistently selected shallow to moderately deep trees (depth 
$\sim$5–6), small learning rates ($\sim$0.02), and 
moderate sub-sampling, while increasing regularization 
and reducing the effective number of boosting 
rounds as the barrier window narrowed --- a 
coherent adjustment that limits complexity when 
overfitting risk decreases.

\subsection{Interpretative analyzes}
\label{sec:3-4}

Beyond the standard fitting predictions enabled by ML 
models, one should not underestimate the importance of
their potential interpretability, so that the emerging 
patterns could, e.g., qualitatively unveil chemical
traits.
Ideally, a model should provide more than a numerical 
relationship extrapolating variations in the target 
variable as a function of the descriptors. 
It should also possess explanatory power, 
allowing to attribute concrete physical and 
hierarchical significance to the observed correlations.

Hence, we shall apply to our results an 
interpretability protocol based on the usual 
implementation of the SHAP method.
Briefly, rooted in Shapley cooperative game theory, 
SHAP transforms complex outputs into additive 
feature contributions, allowing for both local 
and global interpretations of nonlinear and otherwise 
opaque ML models (details of its usage and 
execution can be found in ref. \cite{NIPS2017_8a20a862}).

Next, we only consider XGBoost, as it exhibited 
the best overall performance.
Moreover, we restrict the analysis to the 
better-performing energy regime ($\leq 1.0$ eV), 
comprising a set of 1,495 2D structures.
For each chalcogen (S, Se, and Te), the results are
presented in Fig. \ref{fig:fig6}.
The graphs depict the relative contributions of 
the twenty more relevant predictors --- used in the 
XGBoost model to estimate the energy barriers --- 
ranked in decreasing order of importance, from top 
to bottom.
Each point specifies a given material. 
The horizontal axes gives the SHAP index value, 
gauging the feature influence trend on the prediction.
A positive (negative) index value indicates that the 
feature tends to increase (decrease) the barrier 
energy. 
A predictor has a higher (lower) value if its color
is pinker (bluer). 

\begin{figure*}[t!]
\centering
 \includegraphics[width=1\linewidth]{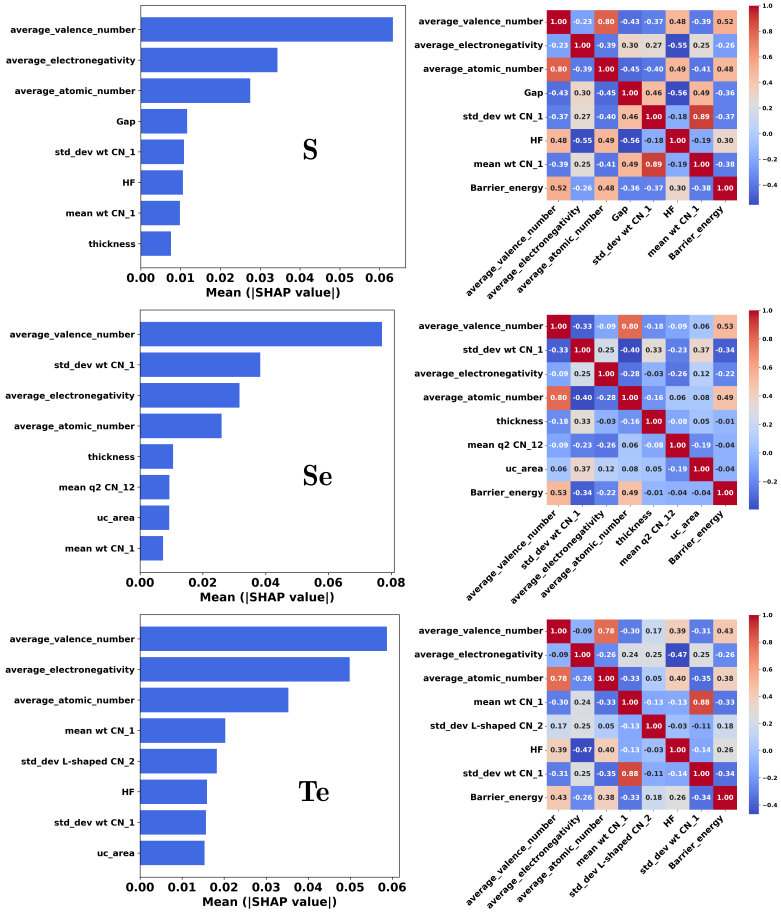}
\caption{The top eight features from the SHAP
analysis (left panels) and the classical Pearson
correlation coefficients from the riginal FSD 
(right panels), for the 2D materials with S, Se 
and Te atom impurity. 
The bar plots rank the most influential descriptors 
according to their mean absolute SHAP value.
The heatmap matrices show the Pearson correlation 
coefficients between the seven most important features
(as identified by SHAP for each atom) and the 
target barrier energy, computed directly from the FSD
without any ML-processing.}
\label{fig:fig7}
\end{figure*}

The relative importance of the top eight descriptors 
for each chalcogen atom are displayed in the left 
panels of Fig. \ref{fig:fig7} --- in units of mean 
modulus SHAP value, horizontal axis.
For the other not shown one hundred twenty four
descriptors, their combined importance in such unity 
is 0.32 (S), 0.37 (Se) and 0.36 (Te).
Below we mostly discuss and interpret the top seven.
Hence, the difference between the seventh and eighth
SHAP index modulus gives an idea of the (somehow
arbitrary) cut-off employed.

The SHAP results show that for the S atom, the top seven
main influential features are:
\begin{itemize}
\item 
\texttt{average\_valence\_number}
\item
\texttt{average\_electronegativity}
\item
\texttt{average\_atomic\_number}
\item
\texttt{Gap} (band gap)
\item 
\texttt{std\_dev\ wt\ CN\_1} 
(weighted standard deviation for 
coordination number 1 atoms)
\item 
\texttt{HF} (heat of formation)
\item
\texttt{mean\ wt\ CN\_1} (weighted standard mean 
of coordination number 1 atoms) 
\end{itemize}
The plots in Fig.~\ref{fig:fig6} reveal that systems
 with lower average valence number and atomic number
and higher electronegativity (observe the colors code)
are more likely to exhibit lower barriers ---
negative SHAP values.
This clearly aligns with S higher chemical affinity 
for more electronegative and compact surfaces. 
Descriptors associated to surfaces with coordination 
number one also play a notable role, suggesting 
(typically) an easier diffusion of S in such structures. 
Although not among the top seven (see the above list
and Fig.~\ref{fig:fig6}),
local topological features, like
\texttt{std\_dev T-shaped CN\_3} (9th) and
\texttt{std\_dev bent 120 degrees CN\_2} (11th) 
are fairly relevant contributors, underscoring 
the influence of surface geometry in  modulating the
energy barrier.

For Se, the seven most relevant descriptors are:
\begin{itemize}
\item
\texttt{average\_valence\_number}
\item
\texttt{std\_dev\ wt\ CN\_1}
\item
\texttt{average\_electronegativity}
\item
\texttt{average\_atomic\_number}
\item 
\texttt{thickness}
\item
\texttt{mean q2 CN\_12}
(mean of the order 2 Steinhardt bond orientation 
parameter for coordination number 12 atoms)
\item 
\texttt{uc\_area} (unit cell area)
\end{itemize}
Among the eight most important descriptors for Se,
six overlap with those for S, although not in the same 
ranking order.
However, Se displays a much higher sensitivity to
the local orientation order variable 
\texttt{mean q2 CN\_12} and the
average valence number feature; compare such a 
feature SHAP values in Fig. \ref{fig:fig7}, which
is around 30\% higher for Se.
This points to a more substantial role of surface 
disorder and irregularities, which is consistent 
with the Se lower electronegativity,
resulting in less localized interactions and thus
a greater dependence on extended regions of the 
surface environment. 
Additionally, thickness and unit cell area are
also important descriptors.
As seen in Fig. \ref{fig:fig6}, as \texttt{thickness} 
decreases and  \texttt{uc\_area} increases (although 
with a certain variability) the energy barrier 
decreases. 
This reinforces the relationship between morphology
and impurity mobility for Se.

For Te, the seven dominant features are:
\begin{itemize}
\item 
\texttt{average\_valence\_number}
\item 
\texttt{average\_electronegativity}
\item 
\texttt{average\_atomic\_number}
\item 
\texttt{mean wt\ CN\_1}
\item 
\texttt{std\_dev\ L-shaped\ CN\_2}
(standard deviation for L-shaped
coordination number 2 atoms)
\item 
\texttt{HF}
\item 
\texttt{std\_dev\ wt\ CN\_1}
\end{itemize}
Similar to Se and S, six of the eight most important
descriptors for Te and S are identical.
Moreover, four of these descriptors also coincide in 
their relative ranking of importance.
This supports the fact that the distributions for
S and Te in Fig. \ref{fig:fig2} are fairly similar.
The Te behavior is predominantly governed by average 
electronic properties, namely, valence, electronegativity, 
atomic number and even \texttt{HF} (6th descriptor), 
with a less pronounced influence from 
coordination-specific features, refer to Fig. 
\ref{fig:fig7}. 
These traits concur with the Te's chemical
characteristics as a heavier, more polarizable, and 
less electronegative element.

More generally, Fig. \ref{fig:fig6} reveals consistent
trends for all the three impurities, while highlighting
distinctive tendencies.
On one hand, the key electronic descriptors, i.e., 
average valence, atomic numbers and electronegativity 
are always among the top four features of S, Se and Te.
Therefore, they form the model core foundation, given 
their strong discriminative power.
On the other hand, S and Se show a balanced sensitivity 
to surface geometry and topology (as coordination type, 
layer thickness, and local disorder), whereas Te is
primarily driven by average electronic properties,
with a weaker influence from local structural 
variations, suggesting a more global and less 
directional interaction with the 2D surfaces.
Also, the moderate relevance of features such as
heat of formation (but less important for Se)
indicates that thermodynamic stability 
and temperature-dependent properties should play 
a certain role in governing impurity diffusion in
2D structures, which is particularly relevant for
applications in catalysis, sensing, and 
2D-material-based devices.

\subsubsection{Pearson correlation of the main
descriptors}

As an additional, independent, assessment of the 
SHAP-derived hierarchy of descriptors --- further
evaluating their predictive power for barrier
energy --- we computed the corresponding Pearson 
correlation (see, e.g., \cite{felippe-2023}) using 
the ``raw'' FSD, i.e., before any ML training.

The correlation matrices for the energy barrier and 
selected descriptors, the collection of seven features 
listed above for each chalcogean atom, are depicted
in the right panels of Fig. \ref{fig:fig7}
(the energy barrier corresponds to the last row,
or equivalently, last column).
The color scale indicates both the sign and 
strength of the relationship.
Obviously, a given pair of descriptors does not 
need to have the same correlation value across 
the distinct chalcogen atom cases due to the way
the correlations are computed \cite{felippe-2023} 
and the collection of variables used.

Comparison between the two types of characterization
(SHAP and Pearson) reveals that the simple calculation 
of linear statistical correlations, aiming to infer 
the most relevant quantities for a target property, 
is a circumscribed procedure.
Indeed, the magnitude of a feature impact on the 
barrier energy as gauged by the ``bare'' correlation
value is not always the same as indicated by SHAP, 
a much more sophisticated and complete nonlinear
interpretabiliy scheme.
Notice that since the Pearson matrices have been
constructed following the same ordering of predictors
by SHAP (left panels), this fact is easy to check by 
inspecting the bottom row of each matrix.
Differently from the SHAP analysis, along it the 
associated coefficients $|c_{ij}|$ (taken in modulus
to comply with the SHAP coefficient, also taken
in modulus) do not monotonically decrease 
(from left to right).

Considering S, $i$ being the \texttt{Barrier\_energy} 
and $j$ being (in decreasing order of 
importance, accordint to SHAP)  
\texttt{average\_valence\_number},
\texttt{average\_electronegativity}, \\ 
\texttt{average\_atomic\_number}, 
\texttt{Gap},
\texttt{std\_dev wt CN\_1}, 
\texttt{HF}, \texttt{mean wt CN\_1}, 
we have, respectively, $|c_{ij}|$ equals to 
0.52, 0.26, 0.48, 0.36, 0.37, 0.30, 0.38.
Note that \texttt{average\_electronegativity}
(\texttt{mean wt CN\_1}), the second (7th)
most important descriptor, has the lowest (third
highest) Pearson correlation, 0.26 (0.38).
So, based just on the raw data, one would conclude 
that \texttt{mean wt CN\_1} is more important than \texttt{average\_electronegativity} to establish
the barrier energy.
Nonetheless, in full agreement with SHAP, 
investigations in the literature point to 
the opposite \cite{gao-2020,lee-2025,shaw-2025}.
Furthermore, ordering discrepancy is likewise observed 
for Se and Te.
In particular, even if we restrict to the three top 
features according to SHAP, they are also classified 
as the three top by the Pearson correlation only for
Se.

The matrices in the right panels of Fig. 
\ref{fig:fig7} also show that the correlations between 
the seven most important predictors have a broad 
dispersion.
While some are well-correlated, others are not.
For the case of S, the top five features 
exhibit a significant degree of mutual correlation, 
with the smallest $|c_{i \neq j}|$ equal to 0.23.
Nonetheless, again considering the top five (even the
top four), the variability is much stronger for Se
and Te, 
with the lowest $|c_{i \neq j}|$'s being 0.03 and 
0.09 for the former and 0.05 and 0.09 for the latter.
These findings may suggest that, when considering the
strength of their reciprocal correlations, there is
no clear pattern for selecting the more pertinent 
predictors.

However, the above observations do not totally
preclude the use of pure statistical measures,
such Pearson's, to help identify characteristics of
appropriate predictors for a model.
As an illustration, from the last row of the
matrices in Fig. \ref{fig:fig7} we can calculate
the standard deviation $\sigma$ of 
$|c_{ \texttt{Barrier\_energy} \ j}|$, with $j$ 
representing the seven most relevant features 
for S, Se and Te.
We find $\sigma_{\mbox{\scriptsize Se}} = 0.220 >
\sigma_{\mbox{\scriptsize S}} = 0.092 >
\sigma_{\mbox{\scriptsize Te}} = 0.084$.
Although not very different for S and Te, these 
$\sigma$'s moderately point to a trend:
the relative importance of the $j$'s is more
evenly (unevenly) distributed in the case of Te
(Se), with S in the middle way.
This qualitatively agrees with the SHAP bar plots 
in Fig. \ref{fig:fig7} and at least partially 
explains the distribution of barrier energies in 
Fig. \ref{fig:fig2}, which is wider for Se, narrower
for S, and even narrower for Te.
In fact, given the great diversity of 2D materials 
analyzed, a higher (lower) variability for the 
influence of the top predictors should lead to 
a bigger (smaller) dispersion of the resulting 
energy barrier, as seen in Fig. \ref{fig:fig2}.

\begin{figure}[t!]
\centering
 \includegraphics[width=1\linewidth]{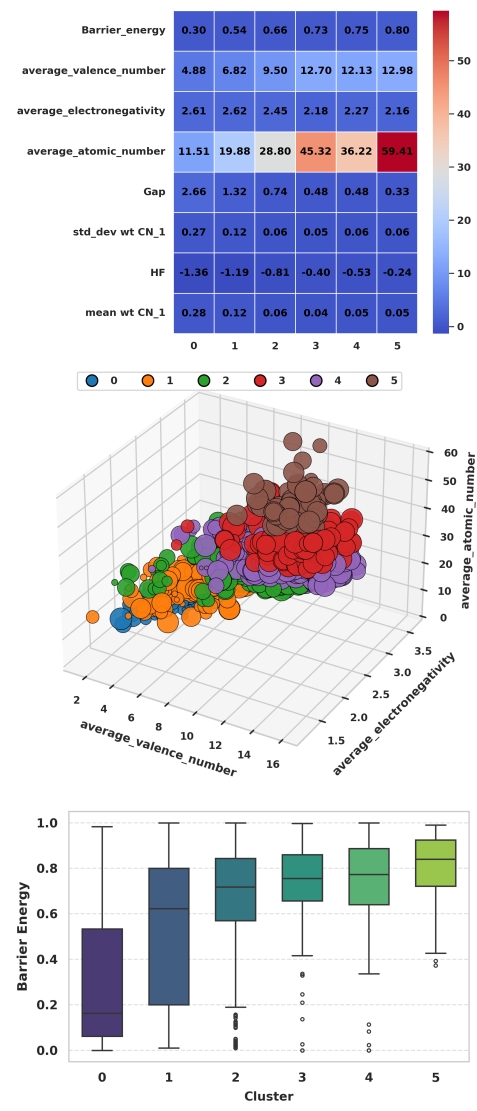}
\caption{For S, materials of Fig. \ref{fig:fig7} 
are classified as cluster 0 to 5 via $K$-means.
Top: mean values of barrier energy and of seven 
relevant predictors (from SHAP) in each cluster. 
Middle: The materials clustering with the 
corresponding values of the three top predictors.
A bullet size indicates the barrier energy value.
Bottom: the barrier energy distribution in each cluster.}
\label{fig:fig8}
\end{figure}

\subsubsection{Classifying the materials:
a cluster approach for S}

The key premise in ML models is that suitably 
chosen descriptors can adequately estimate a target
property, here the barrier energy.
Accordingly, depending on the composition and 
structural characteristics of distinct material
groups --- determining the values of these predictors 
--- the barrier energy is expected to fall within 
specific ranges. 
Conversely, the energies values could reveal systematic
differences in the features distributions, 
providing insight into structure–property 
classification of the systems, useful in
applications.

To explore such possibility, we employ the 
K-means clustering algorithm to divide the materials 
in the database according to their impurity 
mobility driving variables.
Similarly to the Pearson correlations, for an 
independent analyzes we just address the FSD raw 
data, i.e., without any  ML training.
For simplicity, we discuss only S and the 2D systems
for which the barrier energy is not
superior to 1.0 eV (corresponding to 1,495 materials).
We leave Se and Te to the SI.

The K-means is an iterative unsupervised protocol 
that partitions a dataset into collections displaying 
akin attributes.
This is done by calculating either the mean or median
of the data specified properties (in our present case,
those in FSD).
From these quantities, the scheme determines the 
clusters centroids so that distinct points can be 
arranged together into different sets.
The protocol tries to minimizes (maximizes) the variance 
within each cluster (across clusters), so that data 
points in the same (different) cluster(s)  
are as similar (dissimilar) as possible.
For a complete account on the procedure and 
implementation details see, e.g., Ref. \cite{ikotun-2023}.

The results are summarized in Fig. \ref{fig:fig8}.
The method has separated the full collection of 1,495
2D materials into six (the optimal number found) 
distinct clusters, tagged from 0 to 5.
The mean values of the top seven features for S (as
classified by SHAP, see top left panel of Fig. 
\ref{fig:fig7}) as well as of the barrier energy in 
each cluster are presented as a matrix in Fig. 
\ref{fig:fig8}.
The mean barrier energy increases with the cluster label.
In the 3D plot of Fig. \ref{fig:fig8}, each bullet 
represents a specific material, also indicating the 
associated values of the top three descriptors.
While the bullets colors identify which cluster they 
belong to, the diameters match the materials barrier 
energy values.
The bar graph of Fig. \ref{fig:fig8} depicts 
the distribution of barrier energies in each cluster.

Among the clusters, distinct feature patterns are 
discernible.
For example, the barrier energy average value (the 
first row of the matrix in Fig. \ref{fig:fig8}) 
increases through big jumps from clusters 0 to 1
(80.0\%) and 1 to 2 (22.2\%) and then with smaller 
variations from 2 to 3 (10.1\%), 3 to 4 (2.7\%) 
and 4 to 5 (6.7\%) --- 
refer also to the bar chart in Fig. \ref{fig:fig8}.
But this consistently relates to the distribution 
of barrier energy between materials within a cluster.
Indeed, 1 and 0 (in this order) exhibit the greatest 
variations, with 2, 3, 4 and 5 showing reasonably 
similar narrower spread.

\begin{figure}[t!]
\centering
 \includegraphics[width=1.01\linewidth]{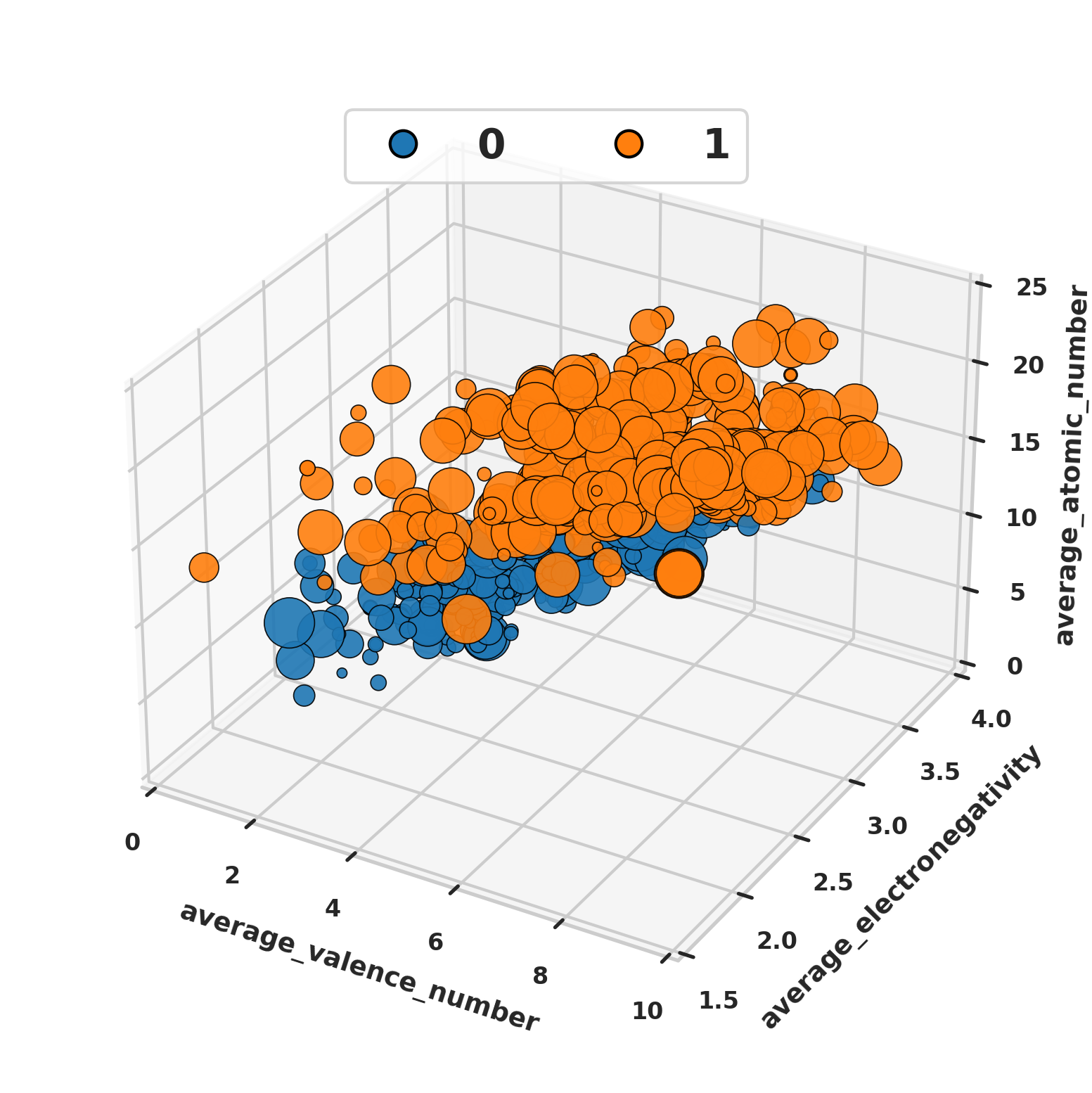}
\caption{The same as in Fig. \ref{fig:fig8},
but showing only clusters 0 and 1.
The plot highlights that materials in these clusters
exhibit a broad distribution of electronegativity.}
\label{fig:fig9}
\end{figure}

As for the average values of the seven primary 
features (matrix in Fig. \ref{fig:fig8}), the
first, third and sixth, namely,
\texttt{average\_valence\_number},
\texttt{average\_atomic\_number} and 
\texttt{HF}, increase with the cluster average 
barrier energy, the exception being for the 
cluster 4.
The fourth descriptor, \texttt{Gap},
systematically decreases, again unless for 
cluster 4.
For the first and third descriptors, this also 
can be verified from the overall trend in the 3D 
plot of Fig. \ref{fig:fig8}.
In particular, notice the broad distribution
of the cluster 4 in the \texttt{average\_valence\_number}
axis and that typically the materials of cluster 3
are above those for cluster 4 along the 
\texttt{average\_atomic\_number} axis.
We recall that these four predictors are highly 
correlated, as indicated by the Pearson matrix for
S in Fig. \ref{fig:fig7}.
The \texttt{average\_electronegativity} (the second 
most important descriptor) exhibits a somehow 
oscillatory trend: 
it remains nearly constant for clusters 0 and 1, 
decreases successively for clusters 2 and 3, and then 
shows a slightly increase followed by a decrease for 
clusters 4 and 5, with a maximum difference of just 
21\% across all clusters.
The fact that the clusters 0 and 1 present the 
highest values of electronegativity is difficult 
to observe from the 3D  plot in Fig. \ref{fig:fig8},
but becomes clear from Fig. \ref{fig:fig9}.
The remaining two main predictors (fifth and seventh), 
associated with \texttt{NC\_1}, exhibit a decaying 
trait toward a steady value as the mean barrier energy
of the clusters increases.

Lastly, we draw a comparison between the 
$K$-means and the SHAP, emphasizing that they are
independent analyses that are not reliant on one 
another (except from selecting which descriptors 
to utilize in the former).
For so, we recall that in Fig. \ref{fig:fig6}, the 
pink (blue) color indicates  higher (lower) values for 
the predictors, and that positive (negative) values of 
the SHAP index points to an increasing (decreasing) 
of the  energy barrier.
Thus, from Fig. \ref{fig:fig6} for S, we clearly
see that in agreement with the $K$-means, the energy 
should increase with the first, second and sixth 
predictors and decrease with the fourth one.
The features associated to \texttt{NC\_1} tend to
decrease the barrier only when they have high values;
if they are small, their influence is less relevant
(note in Fig. \ref{fig:fig6} that the blue-coded 
materials have positive but close to zero SHAP 
index value).
This is in accordance with the clusters 
characterization for these two descriptors.
Finally, as predicted from the $K$-means, the 
somewhat more complex relationship between \texttt{average\_electronegativity} and the barrier 
energy arises from the fact that materials with 
high values for such predictor can either raise or
lower the barrier (see the pink code-color on both
sides of the SHAP axis in Fig. \ref{fig:fig6}
for this feature), while the majority of materials
with lower values (blue code-color) tend to raise
the barrier, albeit mildly.
Hence, SHIP and $K$-means are once more in line.

As a final remark, we observe that clustering 
approaches as the simple and widespread $K$-means 
(but for more sophisticated ones see, e.g., Ref. 
\cite{wosniack-2019}) seem to point to the same 
direction than interpretative protocols, like SHAP.
Nonetheless, the advantage of combining ML with 
interpretative models is to be able to directly 
determine the descriptors relevance order
(refer to Figs. \ref{fig:fig6} and \ref{fig:fig7});
something not possible by just separating the 
materials into cluster classes.

%
%

\section{Final Remarks and Conclusion}
\label{sec:remarks}

Over the past two decades, material informatics has 
emerged as a field that combines AI and data science 
to explore distinct classes of systems.
It focuses on processing information from large 
datasets to discover and characterize novel 
materials. 
With the advent of extensive chemical databases
(e.g., C2DB), the combination of ML models, robust 
statistical methods, new optimization procedures 
and evolutionary algorithms are enabling more 
efficient inverse design and automating  material
analyzes protocols 
\cite{liu-2020,kulik-2022,kanakkithodi-2016}.
Along these lines, we could mention the use of 
various IA strategies for structure forecasting 
and nanoscale distance estimation with the goal of accelerating costly {\em ab initio} methods through integration.
For instance, crystal structure prediction methods 
such as USPEX \cite{glass-2006}, CALYPSO
\cite{wang-2012-calypso} and AIRSS \cite{pickard-2011} 
have been widely employed.
Also, ML platforms have been considered for both 
classification and regression of interatomic distances
(somehow similar to the philosophy in Sec. 
\ref{d-heuristic}), like of C--O in lactone 
derivatives \cite{thi-2023}.

In the present contribution we have fully developed a
novel framework to study a key aspect of atoms diffusion
in 2D systems, the barriers energies.
For so, we have struck into a different direction
than most investigations.
Instead of trying to speed up costly calculations, we 
sought to enhance semi-empirical quantum methods 
--- here HM --- via phenomenologically determining 
effective equilibrium distances (refer to Eq. (\ref{eq:d})); 
moreover combining the obtained results with supervised
ML to improve and speed up the estimation of adsorption 
energy barriers.
In addition, we have used an interpretative protocol 
to gain concrete physical and chemical insight and
ascribe qualitative meaning to the typical 
black-box-like findings of ML approaches.

We should point out that, in contrast to other considerably
more analyzed element classes, such as the already mentioned
alkali metal atoms \cite{Olsson2019,Heidari2023,Ullah2019}, 
here we have chosen to focus on the chalcogen S, Se, and Te 
impurities on 2D structures, for which there are currently insufficiently detailed studies.
Thus, as far as we are aware, this is the most thorough 
investigation presented in the literature.

Employing the EHM (with a kind of mean global 
equilibrium distance $d_{\mbox{\scriptsize eq}}$), we have
computed the energy profiles of S, Se, and Te adsorbed 
on 4,036 systems from the C2DB database. 
From these profiles, average barriers were defined 
and considered as target variables for predictive
ML models. 
A diverse set of physicochemical descriptors was 
extracted using the Matminer library and complemented 
with extra averaged features related to atomic 
properties.
We have tested distinct ML schemes, but determined
that XGBoost consistently outperformed the others in 
accuracy and generalization capacity. 
For the ML investigations,  the original set of 
materials have been reduced to those with energy 
values not superior to 2.0 eV (about 2,560 
materials), since lower values tend 
to attract greater interest in applications 
\cite{han-2015,gao-2015,lee-2025}.

We have verified that further limiting the 
pull of materials to those with energies not 
greater than 1.0 eV (1,495 systems) does increase
the fitting quality and predictability power.
The reason for this is that shifting from a 2.0 
eV window to the more concentrated $\leq$ 1.0 eV 
region stabilizes the feature hierarchy, 
decreasing overfitting and increasing generalization
while providing a dependable and intelligible 
link between migration barriers and atomic-scale 
descriptors.

We have used the SHAP method to characterize the
impact of each structural and electronic quantifier 
on the final energy barriers for the S, Se, and Te
impurities for the materials in the latter energy 
range mentioned above (thus already pertinent for
applications; see Introduction section).
It has confirmed that for XGBoost, the fitting 
correlations are chemically meaningful and consistent 
across impurity types, also capturing the specific
migration physics in each case.
For instance, as expected, descriptors such as average 
valence number, electronegativity, and atomic number 
were always among the most important in determining the
barrier values.
However, topological and geometrical features, such 
as thickness and unit cell area, likewise displayed 
relevant roles.
Regarding morphology, SHAP has shown that overall 
(see Fig. \ref{fig:fig6}) thinner and denser structures 
tend to reduce the migration barrier.

Using S as a case study, independent correlation 
measures have been used to confirm the physical 
significance of the selected features by SHAP.
In particular, Pearson correlation matrices and K-means 
clustering have effectively validated the consistency 
of the results reached by the ML plus interpretability 
scheme.
Nevertheless, SHAP has always being superior to
any statistical inference method.

Importantly, our framework has successfully uncovered
the differences between the chalcogen impurities.
For example, the weighted standard deviation at 
coordination number 1, Fig. \ref{fig:fig7}, shows
that Se is more sensitive to local geometric disorder.  
This suggests that surface imperfections and prolonged 
structural complexity have a greater influence on Se 
migration, which is consistent with its lower 
electronegativity and more diffuse bonding than S.

For Te, the SHAP landscape is broader: the average 
valence number and average electronegativity 
dominate, but local topology descriptors (e.g., 
st. dev. L-shaped CN$_2$ and weighted st. dev. 
CN$_1$) are also relevant, Fig. \ref{fig:fig7}.
Since the Te barrier energy is not too strongly
dependent on just a small number of features, 
we should not expect great dispersion of barrier 
values caused by high variance of only few particular 
predictors across the distinct materials.
This concurs with Fig. \ref{fig:fig2} (note the 
plateau-like behavior for Te) and
with the heavier, more polarizable character and 
reduced sensitivity to subtle surface disorder of Te.

To conclude, we bring up two final points.
First, the current study demonstrates that, when 
combined with reliable ML models, semi-empirical 
approaches provide a workable way to screen adsorption
properties in big databases of 2D materials. 
Moreover, the use of interpretability techniques 
highlights the role of ML not only as an acceleration
tool but also as a way to gain concrete insight 
into the phenomenology of the process investigated.

Second, it must be noted that the current framework 
should not be considered the ultimate step in the classification of 
materials characteristics.
Instead, it constitutes a rather fast selection scheme 
in very large databases.
A second stage of analysis would include detailed 
computations, but then only for the best candidates
for determined applications --- see, e.g., \cite{tian-2024}.
Although along the work we have discussed that 
generally our outcomes are supported by known 
results in the literature, ours are naturally
not {\em ab initio} calculations.
Thence, some discrepancies in actual numerical values 
and even in phenomenological behavior might be the 
case.
For instance, eventually for certain particular 2D materials, instead of diffusion, the atom could 
become ``anchored" on the surface, due to the 
formation of local deep wells, not predicted by the 
protocols here.
Indeed, to determine actual potential profiles we 
should have a position dependent 
$h_{\mbox{\scriptsize eq}}$
instead of an effective $d_{\mbox{\scriptsize eq}}$.
We are presently working on a phenomenological 
$h_{\mbox{\scriptsize eq}}$ based on the idea of
topological DCF descriptors \cite{tromer-2025}, 
which hopefully will be reported in the near future.

In summary, the framework's main benefits are its 
exceptional speed, versatility, and physical interpretability;
being a significant new tool for characterizing the 
diffusion of atoms in vast databases of 2D materials.

\begin{acknowledgement}

The authors acknowledges the following financial support.
MLPJ: The Foundation of the Federal District 
(FAP-DF, grant 00193-00001807/2023-16) and the National 
Council for Scientific and Technological Development 
(CNPq, grant 444921/2024-9).
MGEL: The research grants from CNPq, No. 307512/2023-1 (PQ) 
and No. 04577/2021-0 (Projeto Universal) as well as the
project “Efficiency in uptake, production and distribution
of photovoltaic energy distribution as well as other 
sources of renewable energy sources” 
(No. 88881.311780/2018-00) via CAPES PRINT-UFPR.
PC and MGEL: The 2025 IMI Joint Use Research Program
Workshop (I) 2025b007 on the "Advancing Materials Data,
Design \& Discovery" and I2CNER.
PC is also supported by JSPS KAKENHI Grant-in-Aid for
Scientific Research (C) JP24K06797. PC holds an honorary appointment at La Trobe University and is a member 
of GNAMPA.
ALR thanks CAPES/COFECUB grant $88881.188740/2025-01$ and CNPq under grant numbers $313081/2017-4$, $305335/2020-0$, $309599/2021-0$, and PDPG-FAPDF-CAPES Centro-Oeste grant number $00193-00000867/2024-94$.
MJP thanks FAPERGS, grant $24/2551-0001551-5$, CNPq grants $444431/2024-1$ and $303206/2025-0$
DGS thanks CNPq grant $409792/2024-1$
TASP acknowledges support from the National Institute of Science and Technology on Materials Informatics and CNPq - INCT grant $371610/2023-0$, and FAPEMAT (PRO-2025/00376), CNPq/INCT grant $409174/2024-6$, and PDPG-FAPDF-CAPES Centro-Oeste grant number $00193-00000867/2024-94$.
EAM thanks CNPq, grant number $315324/2023-6$.
CRCR thanks the German Federal Ministry of Education and Research (BMBF) for financial support of the project Innovation-Platform MaterialDigital (\url{www.materialdigital.de}) through project funding FKZ number: 13XP5094A. Part of this work was performed on the HoreKa supercomputer funded by the Ministry of Science, Research, and the Arts Baden-Württemberg and by the Federal Ministry of Education and Research.
ACD thanks FAPDF, grants $00193-00001817/2023-43$ and $00193-00002073/2023-84$, CNPq grant $305174/2023-1$, and PDPG-FAPDF-CAPES Centro-Oeste grant number $00193-00000867/2024-94$.
RMT acknowledges CNPq for a Research Productivity Fellowship (grant no. 307371/2025-5).
All authors acknowledge financial support from CNPq, 
grant 444069/2024-0.
\end{acknowledgement}

\begin{suppinfo}
The Supplementary Information addresses extended results 
for the Se and Te cases, including performance 
metrics and prediction plots for all ML 
models. 
Moreover, a detailed interpretability analysis using 
the SHAP method is also provided.
It  identifies the most influential descriptors for
each impurity. 
All results are based on the same dataset of 
4036 two-dimensional systems from the C2DB database
considered here.
\end{suppinfo}

\bibliography{bibliography}

\end{document}